\documentclass[prl,twocolumn,showpacs,preprintnumbers,amsmath,amssymb,floats]{revtex4}
\usepackage{graphicx}
\usepackage{dcolumn}
\usepackage{bm}

\def\Re{\textrm{Re}}
\def\Im{\textrm{Im}}
\def\be{\begin{equation}}
\def\ee{\end{equation}}
\def\bea{\begin{eqnarray}}
\def\eea{\end{eqnarray}}

\begin{document}
\title{Universality of the single-particle spectra of cuprate superconductors}
\author{Lijun Zhu, Vivek Aji, Arkady Shekhter and C. M. Varma}
\affiliation{Department of Physics and Astronomy, University of California,
Riverside, California 92521}
\begin{abstract}
All the available data for the dispersion and linewidth of the
single-particle spectra above the superconducting gap and the
pseudogap in metallic cuprates for any doping has universal
features. The linewidth is linear in energy below a scale $\omega_c$
and constant above. The cusp in the linewidth at $\omega_c$
mandates, due to causality, a "waterfall", i.e., a vertical feature
in the dispersion. These features are predicted by a recent
microscopic theory. We find that all  data can be quantitatively
fitted by the theory with a coupling constant $\lambda_0$ and an
upper cutoff at $\omega_c$ which vary by less than $50\%$ among the
different cuprates and for varying dopings. The microscopic theory
also gives these values to within factors of $O(2)$.
\end{abstract}
\pacs{74.20.Mn, 74.25.Jb, 74.72.-h}
\maketitle

{\it Introduction}. With increased refinement of technique and
imaginative use, angle-resolved photoemission spectroscopy (ARPES)
on the high temperature superconductors has revealed that novel
physical principles determine the single-particle spectra in such
compounds \cite{campu-review, shen-review}. Recently, the
single-particle spectra over an energy range from the chemical
potential to about 1~eV have been deduced  for various metallic
dopings in different cuprates
\cite{graf1,feng,valla1,graf2,ding,mesot,shen,matrix-elements}.
Also, recently a microscopic theory \cite{aji-cmv} has been
formulated which derives the fluctuations leading to the
phenomenological marginal Fermi liquid(MFL) \cite{mfl} and their
coupling to fermions. MFL had previously been tested only for low
energies and near optimal doping in Bi2212
\cite{valla2,kaminski,abraham-pnas} with adjustable couplings. Here
we test crucial new features of the microscopic theory including its
universality, its cut-off and coupling functions, by comparing with
recent ARPES data in 4 different cuprate families and at different
dopings.

The most important results of these recent ARPES experiments  may be
summarized as follows:

(i) The spectra for energies $\omega$ in the range of interest
(above the superconducting gap and the pseudogap energy scales) are
universal; they have the same functional form for all cuprates and
for all metallic dopings. Moreover, even the parameters in the
functional form vary less than by a factor of $2$ over the entire
range of cuprates for which data is available, irrespective of
whether they are underdoped(UD), optimally doped(OP) or
overdoped(OD).

(ii) The momentum distribution curves (MDC) at constant energy
$\omega$ is a Lorentzian with width $w_{\bf k}(\omega)$.  In the
energy range of interest $w_{\bf k}$ varies linearly with $\omega$
up to a cutoff above which it is approximately a constant. This is
modified if the bare velocity ${\bf v}({\bf k})$ varies within
$w_{\bf k}$, which happens as the bottom of the band is approached.
See Fig.~\ref{fig:mdcwidth} below for representative experimental
data.

(iii) The peak of the MDC as a function of $\omega$ moves with ${\bf
k}$ defining the renormalized dispersion ${\varepsilon}({\bf k})$.
The observed dispersion ${\varepsilon}({\bf k})$ follows the band
structure $\epsilon_{\bf k}$ with a smooth renormalization factor up
to $\omega\approx E_1$. Above $E_1$, the ``velocity"
$d{\varepsilon}({\bf k})/dk$ sharply increases up to another cutoff
$E_2$ where ${\varepsilon}({\bf k})$ resumes the normal dispersion.
The nearly vertical dispersion has been picturesquely termed a ``
waterfall"\cite{graf1}.  In the energy range, $E_1\lesssim \omega
\lesssim E_2$, there is also an indication of multiple
${\varepsilon}({\bf k})$ for fixed $\omega$ \cite{graf2,mesot}.
$E_1$ varies systematically being largest in the $(\pi,\pi)$
direction and smallest in the $(\pi,0)$ direction \cite{mesot}.
Similarly, the position of the ``waterfall" in ${\bf k}$-space
varies systematically.

All these features follow quantitatively from the quantum-critical
fluctuations derived recently \cite{aji-cmv}. We find that, given
the bare band structure $\epsilon_{\bf k}$, all available data can
be fitted with the two parameters of this theory, a sharp cutoff
$\omega_c$ and a coupling constant $\lambda_0$ calculable to factors
of $O(2)$.

{\it Single-particle spectral function}.
The single-particle spectral function deduced by ARPES is given
\be
A({\bf k}, \omega) = {-\Im \Sigma(\omega, {\bf k})/\pi \over [\omega
- \Re\Sigma(\omega,{\bf k})-\epsilon_{\bf k}]^2 + [\Im
\Sigma(\omega,{\bf k})]^2},
\label{eq:spectral}
\ee
where $\Sigma(\omega, {\bf k})$ is the self-energy function. The
band structure $\epsilon_{\bf k}$ is in general fitted by the
tight-binding dispersion\cite{band-structure}.

\begin{figure}[t]
\includegraphics[width=0.7\columnwidth]{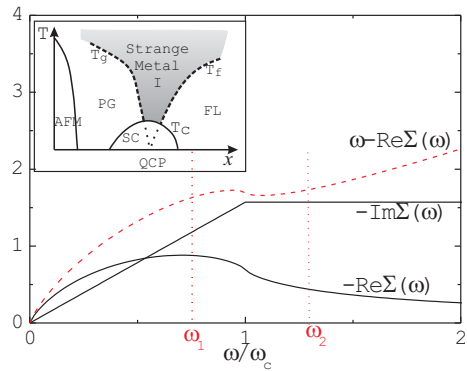}
\caption{The self-energy, and $\omega-\Re\Sigma(\omega)$ as
functions of $\omega$ for $\lambda_0=1$. All quantities are
dimensionless in units of $\omega_c$. The observable dispersion
$\varepsilon({\bf k})$ for a given bare $\epsilon_{\bf k}$, solved
by Eq.~(\ref{eq:varepsilon}) is equivalent to the intersection of
$\omega-\Re\Sigma(\omega)$ with a horizontal line at $\epsilon_{\bf
k}$. As $\omega-\Re\Sigma(\omega)$ has a wide reentrant region for
$\omega_{1} \leq \omega \leq \omega_{2}$, the observed dispersion
falls from $\omega_1$ to $\omega_2$ for a very small variation in
$\epsilon_{\bf k}$ producing the ``waterfall''. Insert shows the
phase-diagram of the cuprates.}
 \label{fig:self-energy}
\end{figure}

{\it Microscopic Theory}. A microscopic theory for the cuprates
\cite{cmv-pr99,cmv-06,aji-cmv} is based on the realization, that the
central organizing feature in the physics of the metallic phase of
the cuprates are quantum critical fluctuations of loop currents. In
this theory the ordering of these loop currents below $T_g$ (PG in
inset of Fig.~\ref{fig:self-energy}) leads to a phase which breaks
time-reversal symmetry but preserves translational symmetry. Direct
evidences for such an ordered state have been obtained by polarized
neutron scattering in YBa$_2$Cu$_3$O$_{6+x}$ \cite{fauque} and by
dichroic ARPES experiments in Bi$_2$Sr$_2$CaCu$_2$O$_{8+x}$
\cite{kaminski02}, for various $x$ in the pseudo-gap phase.  The
properties in the entire funnel shaped region (I) in  inset of
Fig.~\ref{fig:self-energy} are determined by the quantum critical
fluctuations of the loop currents. Therefore universal properties
are predicted for $\omega$ larger than the superconducting gap or
the pseudogap for all $x$ in the metallic phases on either side of
$x_c$.

The microscopic theory of the quantum-critical fluctuations
\cite{aji-cmv} gives their absorptive part to be
\be
\begin{array}{rclc}
\Im\chi({\bf q},\nu)& =&-\chi_0 \tanh{\nu \over 2T}, &  |\nu| < \omega_c;  \\
  && 0, &  |\nu| > \omega_c.
\end{array}
\label{eq:flucspec}
\ee
$\omega_c$ is a cutoff and $\chi_0$ gives the integrated weight of
the fluctuations. In the microscopic theory \cite{aji-cmv},
$\omega_c^2 \approx 2EVR^2$, where $E$ is the  local repulsion or
charging energy parameter, $V$ the nearest neighbor Cu-O
interaction, and $R$ is the dimensionless loop-current order
parameter. $\chi_0 \approx R/\omega_c$. For $E\simeq 5eV$, $V\simeq
1-2eV$ and $R\simeq 0.1$ which is consistent with the neutron
measurements, we expect $\omega_c \approx 0.3-0.5eV$.

{\it Calculation of the Self-energy}. The loop current fluctuations
scatter fermions from ${\bf k}$ to ${\bf k}'$ with the amplitude
$\gamma({\bf k}, {\bf k}')$. From the microscopic model, we find
\cite{footnote2}
\be
\gamma({\bf k}, {\bf k'})=\pm
i\frac{V}{2}(s_x(k+k')s_y(k-k')-x\leftrightarrow y)S_{xy}(k,k'),
\label{eq:coupling}
\ee
where $s_{x,y}(k)\equiv \sin(k_{x,y}a/2)$, $s_{xy}(k) \equiv
\sqrt{s_x^2(k) + s_y^2(k)}$, $S_{xy}(k,k') \equiv
(s_{xy}^{-1}(k)+s_{xy}^{-1}(k'))$. The leading self-energy
contribution is
\be
\Sigma(i\omega_n,{\bf k}) = T\sum_{{\bf q}, i\nu_n} |\gamma({\bf k},
{\bf k+q})|^2 G(i\omega_n+i\nu_n, {\bf k}+{\bf q}) \chi({\bf
q},i\nu_n),
\label{eq:sigma-mfl}
\ee
where $\omega_n$, $\nu_n$ are Matsubara frequencies of the
quasi-particle and the fluctuating mode, respectively. Given a ${\bf
q}$-independent $\chi$ and $\gamma({\bf k}, {\bf k}')$ of the form
of Eq.~(\ref{eq:coupling}), the self-energy variation with ${\bf k}$
on a Fermi surface comes only from the separable $s$-wave part of
$|\gamma({\bf k},{\bf k}')|^2$ which is $\propto (1-\cos k_xa \cos
k_ya)(k \to k')$. This gives $\lambda_k \propto (1-\cos k_xa \cos
k_ya)$ which varies by about a factor of 2 from the $(\pi,\pi)$ to
the $(\pi,0)$ directions for the Fermi surface of Bi2212 near
optimal doping.

At $T=0$ the self-energy is easily evaluated to be
\begin{align}
\Im\Sigma(\omega, {\bf k}) =& -\lambda({\bf k}) \frac{\pi}2
\begin{cases} |\omega|,& |\omega|<\omega_c \\ \omega_c,&
|\omega|>\omega_c \end{cases}
\notag\\
\Re\Sigma(\omega,{\bf k}) = &-{\lambda({\bf k}) \over 2}
\Big[\omega\ln\frac{\omega_c}{|\omega|}
+(\omega-\omega_c)\ln\frac{|\omega-\omega_c|}{\omega_c} \notag \\
 & -
(\omega\to -\omega) \Big],
\label{eq:self-energy}
\end{align}
where $\lambda({\bf k}) = \lambda_0 \langle \gamma^2 \rangle_{k'}$;
$\lambda_0=N(0) (V^2/4)\chi_0$ and $\langle \gamma^2\rangle_{k'}$ is
the average of $|\gamma({\bf k}, {\bf k}')|^2$ over ${\bf k}'$ on
the Fermi surface. For the density of states per one spin species
$N(0) \approx 1(eV)^{-1}$ and other parameters used above, we expect
$\lambda_0 \approx 1$.

Given such a weakly momentum dependent self-energy, the vertex
corrections \cite{migdal} to the self-energy are only of
$O(\lambda\omega_c/W)$, where $W$ is the bare bandwidth of the
conduction band. Using the $\omega_c$ and $\lambda$ fitted to the
experiments, this ratio is of $O(0.1)$. The remaining processes,
repeated scattering (self-consistent Born approximation) produce no
singular corrections. At low energies compared to $\omega_c$,
Eq.~(\ref{eq:self-energy}) reduces to the MFL form deduced earlier
\cite{mfl}, except for the weak momentum dependence.

Given the sharp change of the slope in the imaginary part near
$\omega_c$, the real part has a logarithmic divergence in its slope
at $\omega_c$ before changing from its low energy form $\propto
\omega\log|\omega|$ to $1/\omega$ for $\omega \gg \omega_c$. This
sharp variation of $\Re\Sigma(\omega)$ near $\omega_c$ is
responsible for the observed ``waterfall'' feature as we now proceed
to show.

\begin{figure}[tbh]
\includegraphics[width=0.8\columnwidth]{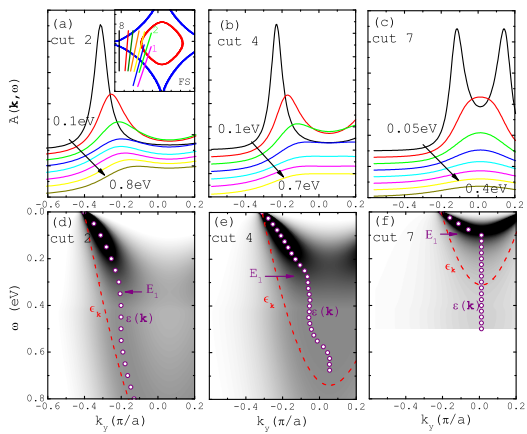}
\caption{(a),(b) and (c): calculated MDCs for three of the momentum
cuts 2, 4 and 7 shown in the inset to Fig.~(a) for which data is
available from Ref.~\onlinecite{mesot}. The MDC's are shown at
various energies $\omega$ labeled in the figures. Figs.~(d),(e) and
(f) are spectral intensity maps, for the same cuts correspondingly.
From the intensity maps, we can identify the dispersion
$\varepsilon({\bf k})$ marked by circles; the bare dispersion
$\epsilon_{\bf k}$ are shown in dashed lines. The inset of Fig.~(a)
shows the Fermi surface and eight momentum cuts done in experiments
\cite{mesot}.  It also shows the positions where the ``waterfall''
are expected in ${\bf k}$-space for radial cuts with the additional
contour drawn inside the Fermi surface.}
\label{fig:mdc}
\end{figure}

\begin{figure}[tbh]
\includegraphics[width=0.7\columnwidth]{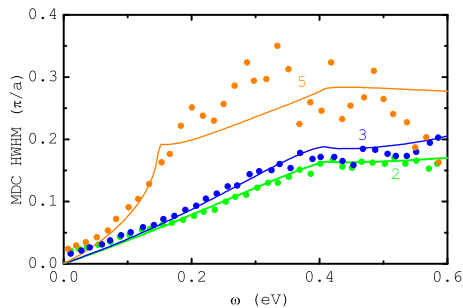}
\caption{The MDC half-width at half-maximum $w_{\bf k}(\omega)$ is
shown for the cuts 2 ,3 and 5 of the inset of Fig.~\ref{fig:mdc}(a).
The experimental data for the same cuts from Fig.~2 of
Ref.~\onlinecite{mesot} is also shown. Note that the experiments
quote are done with an energy resolution of $30$ meV, which accounts
for the deviation from the theory at low energies. Higher resolution
data \cite{kaminski} confined to lower energies is consistent with
the theory.}
\label{fig:mdcwidth}
\end{figure}

{\it The Waterfall}.
The dispersion of the quasi-particles, ${\varepsilon}({\bf k})$ given by
\be
{\varepsilon}({\bf k})
-\Re\Sigma({\varepsilon}({\bf k}))-\epsilon_{\bf k}=0.
\label{eq:varepsilon}
\ee
As shown in Fig.~\ref{fig:self-energy}, $\omega-\Re\Sigma(\omega)$
has a wide reentrant region from $\omega_{1}\leq\omega\leq
\omega_{2}$. The solution of Eq.~(\ref{eq:varepsilon}) therefore
produces a ``waterfall" in the dispersion ${\varepsilon}({\bf k})$
because it varies over the large energy range $\omega_1$ to
$\omega_2$ for a very small variation in ${\bf k}$. The multiple
solutions obtained in this region are within $\Im\Sigma(\omega)$ for
$\lambda$ of O(1). Above $\omega \simeq\omega_2$, the dispersion
becomes just a renormalized band structure. The calculated
``waterfall'' is shown in Fig.~\ref{fig:mdc}. The spectral intensity
maps in Fig.~\ref{fig:mdc}(d-f) should be compared with Fig.~1(a-c)
of Ref.~[\onlinecite{mesot}].

\begin{figure}[tbh]
\includegraphics[width=0.7\columnwidth]{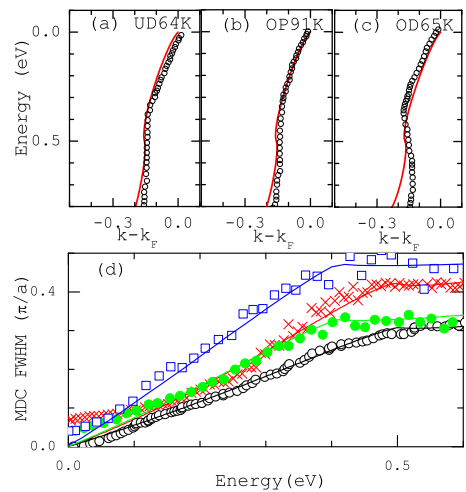}
\caption{Comparison between experimental and theory results
(represented by symbols and lines, respectively) for various cuprate
samples. (a)-(c) are calculated dispersions for three Pb-doped
Bi2212 samples along the nodal cuts: UD with $T_c=64K$, OP with
$T_c=91K$ and OD with $T_c=65K$. The experimental data shown are
extracted from Fig. 1 of Ref.~\cite{graf1}. The tight-binding
fitting parameters of the band structure are taken from
Ref.~\cite{kordyuk2003}. All these samples are fitted by the
parameters $\omega_c=0.5$eV(for all) and $\lambda_0=0.98$, $1.01$,
and $1.05$, respectively. (d) shows the MDC linewidths (full width
at half maximum) for different cuprate samples. $\circ$, $\times$,
$\bullet$ and $\square$ represent OP-Bi2201 (nodal cut,
Ref.\cite{shen}), OP-Bi2212 (nodal cut, Ref.~\cite{graf-comm}), LSCO
0.17 (cut 2 in Fig. 2 of Ref.~\cite{mesot}) and LSCO 0.145 (cut 1 in
Fig. 3 of Ref.~\cite{mesot2}), respectively. The corresponding
theory fitting parameters are: $\lambda_0=0.99$, $\omega_c=0.5$eV;
$\lambda_0=1.01$, $\omega_c$=0.5eV; $\lambda_0=1.09$,
$\omega_c$=0.41eV and $\lambda_0=1.64$, $\omega_c$=0.41eV. }
\label{fig:diffsamp}
\end{figure}

{\it Comparison with Experiments}: The calculated  self-energy at
$\lambda_0 = 1$ (suitable to fit the experimental data \cite{mesot}
for La$_{1.83}$Sr$_{0.17}$CuO$_4$) is shown in
Fig.~\ref{fig:self-energy}. The experimental MDC width for this
compound and the calculated widths for three different cuts are
compared with experiment in Fig.\ref{fig:mdcwidth}. In
Fig.\ref{fig:diffsamp}(a)-(c), we compare the experiments
\cite{graf1} for the dispersion of three Bi2212 samples at different
dopings with calculations with $\omega_c =0.5, \lambda_0 \sim 1$. In
Fig.\ref{fig:diffsamp}(d), we compare the measured linewidth for an
UD-LSCO sample, an OP-Bi2201 sample and an OP-Bi2212 sample with
calculations with parameters given in the figure caption.

{\it Universality of the Data}: The data and the comparison with
experiments in Fig.\ref{fig:mdcwidth} and Fig.
\ref{fig:diffsamp}(a)-(d) attest to the universality of the
single-particle spectra of the cuprates and of the quantitative
success of the theory.  Now we consider in detail each of the points
(i) to (iii) of the experimental data and explain them successively.

(i) The physical properties in any quantum critical regime are
universal, controlled by the scale-invariant critical fluctuations.
Specifically, for $\omega$ larger than the superconducting gap or
the pseudogap the self-energy is of MFL form and given in terms of
only the two parameters $\omega_c$, $\lambda_0$ for each compound
for all $x$. Weak dependencies in these parameters from variation in
microscopic parameters due to varying $x$ or $T$ may occur of
course. We find however that for a given compound, a single value of
these parameters is adequate to fit all the available data for
different $x$ and for all momentum directions.

It is worth noting that the spectra for energies below the pseudogap
energy and $T\leq{T}_g$ is also scale-invariant with a new scale
$\propto{T}_g(x)$ \cite{kanigel,lijun-cmv}.

(ii) Suppose at certain energy $\omega$, Eq.~(\ref{eq:varepsilon})
is satisfied for ${\bf k} = {\bf k}_0$. Since the self-energy does
not depend significantly on ${\bf k}$, we can expand the spectral
function in $({\bf k}-{\bf k}_0)$. The MDC is then a Lorentzian with
width $w_{\bf k}$ given by $\Im\Sigma(\omega)/{v}({\bf k}_0)$ where
$v({\bf k}_0) = v_y({\bf k}_0) + v_x({\bf k}_0)
(k_x-k_{x0})/(k_y-k_{y0})$, is the bare velocity in the momentum-cut
direction. This expansion also requires that within $({\bf k}-{\bf
k}_0) \approx w_{\bf k}$, the velocity ${\bf v}_{\bf k}$ is nearly a
constant.

As discussed above $\Im\Sigma(\omega)$ increases linearly in
$\omega$ for $\omega \lesssim \omega_c$ and is constant beyond.
Therefore if ${\bf v}_0({\bf k})$ varies slowly with ${\bf k}$ as in
cut 2 in Fig.~\ref{fig:mdc}, MDC linewidths also vary linearly in
$\omega$, i.e., $w_{\bf k} \propto \omega$.  Away from the nodal
momentum directions, ${\bf v}_0({\bf k})$ varies considerably as in
cut 4 and higher of Fig.~\ref{fig:mdc}. As a result, MDCs' linewidth
deviates from the linear-$\omega$ dependence. This accounts for the
MDC width of cut 5 shown as an example in Fig.~\ref{fig:mdcwidth}
and the higher cuts. If the MDC linewidth is multiplied by the bare
velocity at each ${\bf k}$ in any direction, a linear dependence of
the width with $\omega$ is obtained both in theory and the
experiments.

(iii) Comparing Figs.~\ref{fig:mdc}(d-f),  we can see that there are
two distinct reasons for the ``waterfalls''.  If $\epsilon_{\bf k}$
reaches $\omega_1 -\Re \Sigma(\omega_1)$ at ${\bf k} \approx {\bf
k}_0$ as ${\bf k}$ is varied along the momentum cut, e.g., cut 2 in
Fig.~\ref{fig:mdc}, ${\varepsilon}({\bf k})$ follows the
``waterfall'' between $\omega_1$ and $\omega_2$, which correspond to
$E_1$ and $E_2$ defined in experiments.

If the momentum cuts are sufficiently away from the nodal cut such
that the bottom of the band is very shallow, $\epsilon_{\bf k}$
never reaches $\omega_1 -\Re \Sigma(\omega_1)$; e.g., cuts 5-8 in
Fig.~\ref{fig:mdc}. The observed dispersion ${\varepsilon}({\bf k})$
then follows Eq.~(\ref{eq:varepsilon}) to its maximum value at the
bottom of the band ${\bf k}_m$. For higher energies, there are no
solutions to Eq.~(\ref{eq:varepsilon}). In this case the MDC curves
stay centered at ${\bf k}_m$ which leads to another type of
``waterfall''.  $E_1$ in this case is nearly the energy of the
bottom of the renormalized band, and gets continuously smaller as
the bottom of the band (where the velocity is zero) becomes
continuously more shallow from the $(\pi,\pi)$ to the $(\pi,0)$
direction. The variation of the position of the ``waterfall''s,
Fig.~3 of Ref.~[\onlinecite{mesot}] and Fig.~3 of
Ref.~[\onlinecite{graf2}] is thereby explained. In addition, the
linewidth is no longer given by $\Im\Sigma/v_0$, leading to an
additional cusp in linewidth at $E_1$ (e.g., cut 5 in
Fig.~\ref{fig:mdcwidth}).

However, if radial cuts are taken to avoid the shallow band, the
position of the ``waterfall'' in momentum space is always the locus
of ${\bf k}$ where ${\varepsilon}({\bf k}) \approx \omega_1$. This
locus is shown in Fig.~\ref{fig:mdc}(a) for radial cuts and is to be
compared with data in Fig.~4  of Ref.~[\onlinecite{mesot}] and
Fig.~4 of Ref.~[\onlinecite{graf2}].

{\it Concluding remarks}. The experimental results discussed place
strong constraints on a theory applicable to the cuprates.
Specifically, the experiments give a scattering rate linear in
$\omega$ up to a sharp cut-off at $\omega_c$ and constant above with
a coefficient which is a weak function of ${\bf k}$. This behavior
is found in the entire 'strange metal' region of the phase diagram.
We do not know any ideas proposed for cuprates besides those
discussed here which give these properties.

In this paper, we have pointed out the universal aspects of the measured
single-particle self-energy in cuprates and shown that its
functional form and even its magnitude is consistent with the recent
microscopic theory of quantum critical fluctuations \cite{aji-cmv}.
These fluctuations are predicated on the existence of an unusual
symmetry breaking in underdoped cuprates for which considerable
experimental evidence has also been adduced.

{\it Acknowledgments}. CMV is especially grateful to J. Mesot for
interesting him in the problem and to him, J. Chang, S. Paillh\`{e}s
and C. Mudry for a detailed discussion of the data. Thanks are also
due to J. Graf and A. Lanzara for communications regarding their
data.

\end{document}